\documentclass[amsmath,amssymb,aps]{revtex4-2}

\usepackage{graphicx}
\usepackage{dcolumn}
\usepackage{bm}
\usepackage{dsfont}
\usepackage{natbib}
\usepackage{graphicx,amsmath}
\usepackage{epsf}

\begin{document}

\title{The Formation of Photon-Molecules in Nanoscale Waveguides}
\author{Hashem Zoubi}
\email{hashemz@hit.ac.il}
\affiliation{Department of Physics, Holon Institute of Technology, Holon 5810201, Israel}
\date{21 July, 2021}

\begin{abstract}
We investigate the formation of photon bound states in a system of interacting photons inside nanoscale wires. The photons interact through the exchange of vibrational modes induced along the waveguide mainly due to radiation pressure. The problem of many-body photons is treated in using the formalism of contour Green's functions under the T-matrix approximation. The complex pole of the T-matrix is a signature for the appearance of photon-molecules. The analysis of such singularity provides the critical temperature at which the T-matrix approximation breaks down and photon-molecules appear. For strongly interacting slow photons the amplitude of the photon-molecule wavefunction acquires a significant quantum nonlinear phase inside the nanowire. Photon bound-states can be implemented for quantum information processing as quantum logic gates, e.g. for $\pi$ phase shift the photon-molecule is shown to serve as a Z-controlled gate.
\end{abstract}

\maketitle

\section{Introduction}

The noninteracting nature of photons makes them efficient for long distance communication but nonefficient for information processing. The interaction
among photons is the corner stone for their physical implementation in
quantum information processing and quantum logic gates \cite{OBrien2007}. Quantum nonlinear
optics involving single photons is a recent and hot topic with
importance for fundamental physics and applications \cite{Chang2014}, e.g., for
photonic switches, optical modulation and manipulation, the generation of
single photons on demand, nonlinear spectroscopy, memory devices
and transistors, with impact on physical and biological
sciences \cite{Phillips2001,Miller2010,Chen2013,Reiserer2015}. On the other
hand, in classical
nonlinear optics propagating light in a medium can modify the optical properties of the
material, e.g., by producing intensity dependent refractive index, which implies
powerful lasers. Therefore, conventional nonlinear
optics found to be negligible at the level of individual photons \cite{Boyd2008,Agrawal2013}, and hence
the need for efficient new mechanisms is appealing.

Remarkable advances in the
search for the realization of strong interactions among single photons have
been accumulated in the last decades. Cavity Quantum
Electrodynamics (QED) have been among the first experiments for achieving effective
photon-photon interactions by enhancing the light-matter coupling in
localizing atoms inside high finesse cavities
\cite{Turchette1995,Birnbaum2005,Haroche2006,Reiserer2015}. Cooling and trapping single atoms within a cavity is a complex mission, hence
cavity QED experiments have been
successfully extended into solid state systems including quantum dots in semiconductors \cite{Michler2000,Pelton2002,Fushman2008,Lodahl2015} or
nitrogen-vacancy centers in diamond \cite{Faraon2011}. However, in a recent
successful experiment a deterministic photon-photon
quantum gate has been realized for a single atom in an optical resonator
\cite{Hacker2016,Welte2018}. Even though, the discreteness of the cavity
spectrum gives rise to output photons with narrow bands,
besides dephasing and decay effects of the excited electronic states, which put limitations on the
quantum nonlinear optics performance.

In order to overcome the limitations imposed by confining the photons in high
quality optical cavity, the search has been turned to strong
atom-photon coupling in cavity free
environment \cite{Tey2008,Hammerer2010,Volz2014,Prasad2020}. For example, in using Rydberg atoms in
a dense medium \cite{Gorshkov2011,Peyronel2012,Firstenberg2013,Firstenberg2016,Thompson2017}
by exploiting Rydberg blockade phenomena \cite{Lukin2001} in the Electromagnetic Induced
Transparency (EIT) scheme \cite{Harris1990,Fleischhauer2005}. In such systems the significant enhancement of photon-photon
interactions is mainly due to the achievement of slow light using EIT environment with
extremely narrow transparency band \cite{Petrosyan2011}. Slowing
and stopping the light has been observed in cold and ultracold boson gases \cite{Phillips2001,Hau2008,Pritchard2010}. The interaction between photons is exploited to demonstrate a photon–photon quantum gate, extending the potential of Rydberg systems as a platform for quantum communication and quantum networking \cite{Tiarks2019}.

In parallel, solid-state set-ups of optical fibers \cite{Zhu2007,Thevenaz2008,Douglas2015,Goban2015} and photonic
crystals \cite{Russell2006,Baba2008,Eichenfield2009} have received significant interest,
as they can be easily integrated into all-optical on-chip platforms. In
particular optical fibers can realize tunable delays of optical signals with
the possibility of achieving fast and slow light in a comparatively wide
bandwidth \cite{Okawachi2005,Song2005,Herraez2006}. The most efficient
nonlinear process inside optical fibers is Stimulated Brillouin Scattering (SBS), that is the scattering of optical photons by long lived
acoustic phonons commonly induced by electrostriction
\cite{Kim2015}. Recent progress in the fabrication of nanoscale waveguides, in
which the wavelength of the light becomes larger than the waveguide
dimension, achieved a breakthrough in SBS \cite{Pant2011,Shin2013,Eggleton2013}. In this regime the coupling of photons and phonons is significantly
enhanced due to radiation pressure dominating over electrostriction
\cite{Rakich2012,VanLaer2015a,Zoubi2016} with significant implications for the field of
continuum quantum optomechanics \cite{Rakich2016,Zoubi2018,Zoubi2019,Zoubi2020}.

We have explored the possibility of achieving a significant nonlinear phase shift among
photons propagating in nanoscale waveguides. The interaction among photons is
mediated by vibrational modes and induced through SBS, where an effective photon-photon interaction Hamiltonian is
derived \cite{Zoubi2017}. Moreover, we
have introduced a configuration for slowing down photons by several orders of
magnitude via SBS involving sound waves and pump fields. We extracted the
conditions for maintaining vanishing amplitude gain or loss for slowly
propagating photons while keeping the influence of thermal phonons to the
minimum.

In the present paper we search for the possibility of the formation of exotic photon molecules. Such issue has been addressed
before in the context of an atomic ensemble with Rydberg blockade where attractive photon-photon interactions can be achieved \cite{Firstenberg2013,Guerreiro2014,Maghrebi2015}. Here we aim to
examine the possibility of the formation of two-photon bound states that is induced through the exchange of phonons. To this end we exploit our previous results where effective photon-photon coupling have been derived and that found to be attractive or repulsive \cite{Zoubi2016,Zoubi2017}. We use the tool of contour Green's functions and we derive a hierarchy of equations that can be truncated by the appeal to the T-matrix approximation \cite{Kadanoff1962,Abrikosov1963,Fetter1971,Mahan2000,Stefanucci2013}. The T-matrix approximation allows us to treat the scattering of particles in many-body systems, that is in a medium of interacting particles. The breakdown of the T-matrix approximation, which appears as a singularity in the solution, indicates the formation of bound states. We concentrate in the case of slow photons propagating in one-dimensional nanoscale wires, and we look for the critical temperature at which photon molecules form. The interactions among two counter-propagating photons give rise to an accumulated quantum nonlinear phase, and we show how such phase can be exploited in order to implement photon molecules as quantum logic gates.

After presenting the Hamiltonian of interacting photons in section 2, we introduce the contour Green's functions in the complex plane, where the functions allow considering quantum and ensemble averages on equal footing. In section 3 we derive the equations of motion and solve them in applying the T-matrix approximation. We treat the one-dimensional case where a complex pole appears in the T-matrix that represents the appearance of photon molecules. The photon bound states are discussed in section 4, and the equation of motion of the state amplitude is solved to yield a quantum nonlinear phase that shown to be useful for achieving quantum logic gates. The conclusion is given in section 5. In the appendix we analyze the properties of two-point correlators in Keldysh space.

\section{Interacting photons}

We consider propagating optical photons in nanoscale waveguides (see Fig.1), where the photons can propagate to the left or to the right with effective group velocity $v_e$. In our previous work we introduced a configuration for controlling the group velocity and to achieve relatively slow photons by exploiting the coupling between photons and acoustic phonons that assisted by additional pump fields \cite{Zoubi2017}. The photons are shown to propagate without gain or loss along the waveguide with a linear dispersion of $\omega_k=\omega_0+v_ek$ in the appropriate region with wavenumber $k$, and $\omega_0$ appears due to the transverse nanoscale confinement of dimension in the range of hundred nanometers, (as seen in Fig.2). Slow photons are necessary for achieving manifest phenomena of nonlinear quantum optics in waveguides of several centimeter length \cite{Zoubi2016}. The photons found to interact by exchanging optical phonons (vibrational modes), that is the mechanism for achieving photon bound states, and which is the main concern in the present paper. The effective photon-photon coupling, $v$, is detunable and can be positive or negative that indicates a repulsive or an attractive interaction among the photons \cite{Zoubi2017}. In typical nanoscale waveguides one can achieve photon-photon coupling per meter of about $1$~MHz, and effective group velocity of $10^5$~m/s, as have been shown by us in \cite{Zoubi2017}.

\begin{figure}
\includegraphics[width=0.4\linewidth]{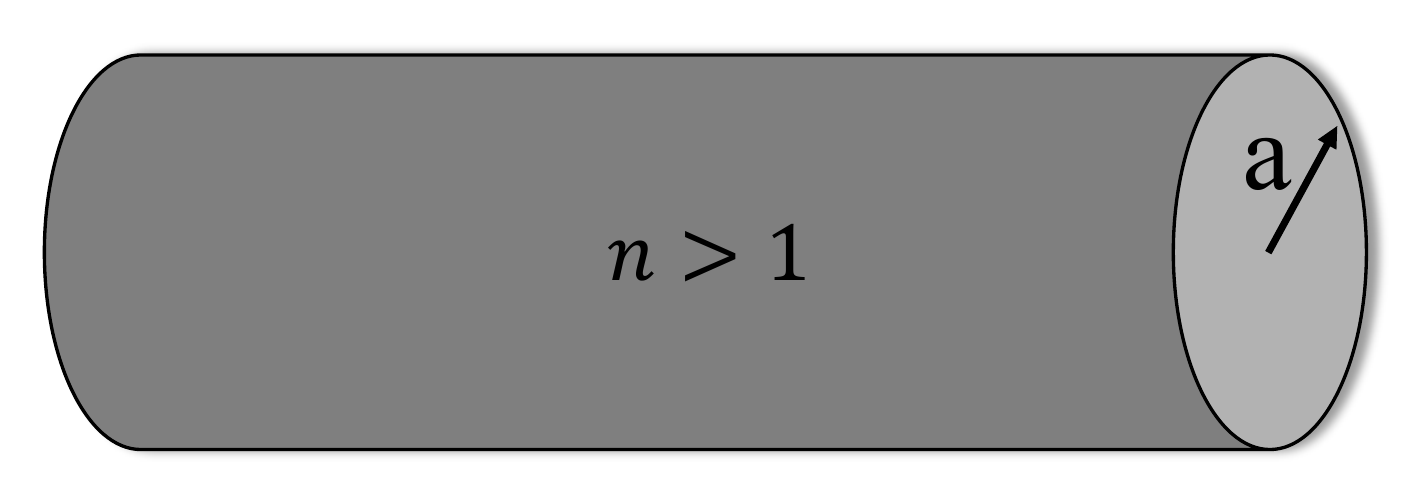}
\caption{The nanoscale wire of radius $a$ of several hundreds of nanometers and length of several centimeters. The fiber made of dielectric material with refractive index $n$ larger than that of the surrounding air.}
\label{PhotPhonDis}
\end{figure}

The total Hamiltonian is given by $\hat{H}=\hat{H}_0+\hat{H}_I$. The free part Hamiltonian in real space is
\begin{equation}
\hat{H}_0=\int d{\bf x}d{\bf x}' \hat\psi^{\dagger}({\bf x})\langle {\bf x}|\hat{h}|{\bf x}'\rangle\hat\psi({\bf x}'),
\end{equation}
where $\langle {\bf x}|\hat{h}|{\bf x}'\rangle=\delta({\bf x}-{\bf x}')h({\bf x},-i{\bf \nabla})$, and the interaction  part Hamiltonian is
\begin{equation}
\hat{H}_I=\frac{1}{2}\int d{\bf x}d{\bf x}'~v({\bf x},{\bf x}')~\hat\psi^{\dagger}({\bf x})\hat\psi^{\dagger}({\bf x}')\hat\psi({\bf x}')\hat\psi({\bf x}),
\end{equation}
where the interaction potential obeys $v({\bf x},{\bf x}')=v({\bf x}',{\bf x})$. Here $\hat\psi({\bf x})$ and $\hat\psi^{\dagger}({\bf x})$ are the field annihilation and creation operators, respectively. For bosons the operators obey the commutation relations
\begin{equation}
\left[\hat\psi({\bf x}), \hat\psi({\bf y})\right]=\left[\hat\psi^{\dagger}({\bf x}), \hat\psi^{\dagger}({\bf y})\right]=0,\ \ \ \left[\hat\psi({\bf x}), \hat\psi^{\dagger}({\bf y})\right]=\delta({\bf x}-{\bf y}).
\end{equation}
For linear dispersion we get $\langle {\bf x}|\hat{h}|{\bf x}'\rangle=\delta({\bf x}-{\bf x}')(\omega_0-iv_e{\bf \nabla})$, and the local interaction potential is 
$v({\bf x},{\bf x}')=v\delta({\bf x}-{\bf x}')$.

\begin{figure}
\includegraphics[width=0.4\linewidth]{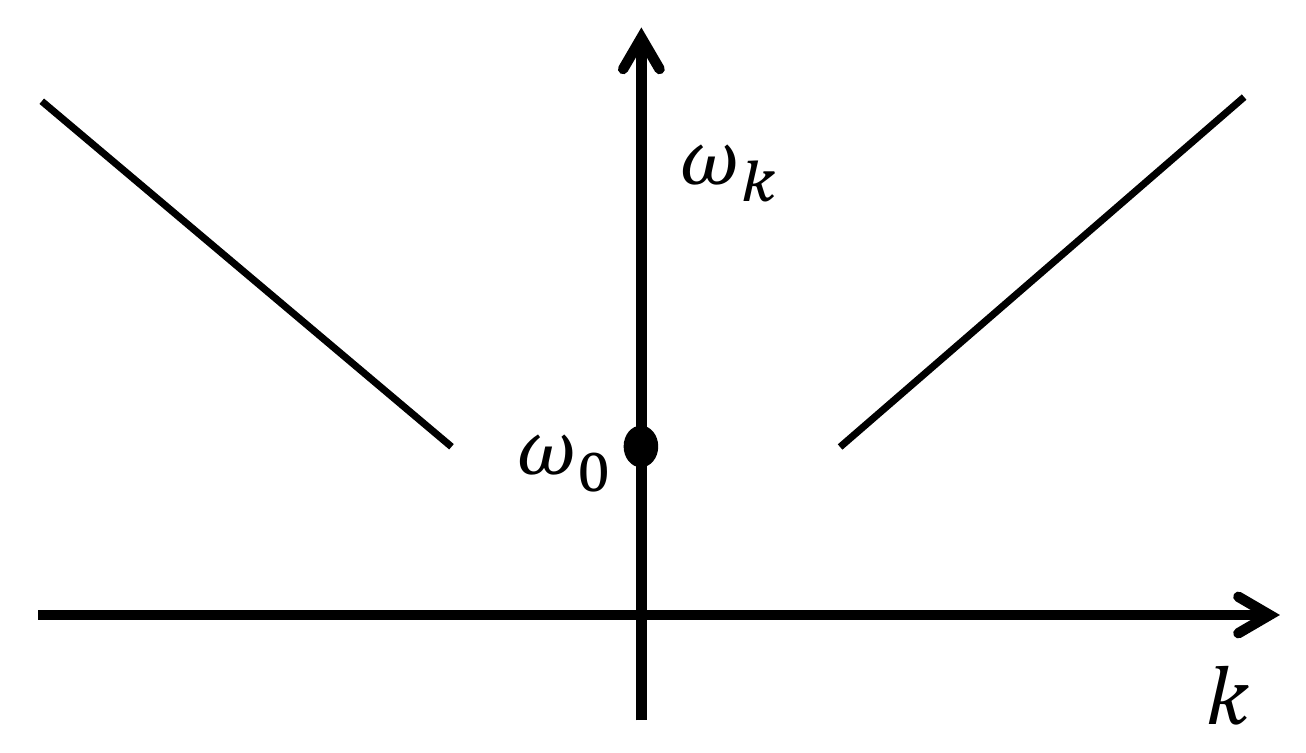}
\caption{The photon dispersion $\omega_k$ as a function of wavenumber $k$, that includes two branches of photons propagating to the left and the right of the nanoscale waveguide. Here $\omega_0$ is a minimum frequency appears due to transverse confinement.}
\label{PhotPhonDis}
\end{figure}

\subsection{The Contour Green's Functions}

The many-particle system of interacting photons can be treated in using the tool of Green's functions \cite{Kadanoff1962,Abrikosov1963,Fetter1971,Mahan2000,Stefanucci2013}. We adopt a unified framework of the contour formalism that allows us treating time-dependent problems and statistical averages at finite temperature of thermal equilibrium. The contour n-particle Green's function is defined by \cite{Stefanucci2013}
\begin{equation}\label{nGreen}
G_n(1,\cdots,n;1',\cdots,n')=\frac{1}{i^n}\frac{ \text{Tr} \left[ {\cal T} \left\{ e^{-i\int_{\gamma}d\bar{z}\hat{H}(\bar{z}) } \hat\psi_H(1)\cdots\hat\psi_H(n)\hat\psi_H^{\dagger}(n')\cdots\hat\psi_H^{\dagger}(1') \right\} \right] } {\text{Tr} \left[ {\cal T} \left\{ e^{-i\int_{\gamma}d\bar{z}\hat{H}(\bar{z}) } \right\}\right] }.
\end{equation}
The integral is along the contour $\gamma$ that appears in figure (3) in the complex plane, where a general point on the contour is denoted by $z$. The horizontal part contains the forward branch $\gamma_-$ from time $t_0$ up to time $t$, and the backward branch $\gamma_+$ from $t$ back to $t_0$, along the real time axis. Then,  a point on branch $\gamma_-$ at time $t'$ is denoted by $z'=t'_-$ and on branch $\gamma_+$ by $z'=t'_+$. The horizontal part is extended to infinity without affecting the result and is called the Schwinger-Keldysh contour. The vertical part $\gamma_M$ of the contour starts at $z_a=t_0$ and end at $z_b=t_0-i\beta$ along the imaginary axis, with the constrain $z_b-z_a=i\beta$, where at temperature $T$ we have $\beta=\frac{1}{k_BT}$.

The field operators $\psi_H(j)$ and $\hat\psi_H^{\dagger}(j)$ are in the contour Heisenberg picture, where we used the short notation $(j\equiv{\bf x}_j,z_j)$. For an operator $\hat{O}(z)$ with explicit time dependent in the contour Heisenberg picture we have $\hat{O}_H(z)=\hat{U}(z_\text{i},z)\hat{O}(z)\hat{U}(z,z_\text{i})$, where $z_\text{i}$ is the initial point of the contour, (and $z_\text{f}$ is the final point). The contour evolution operator, for $z_2$ later than $z_1$, is defined by
\begin{equation}
\hat{U}(z_2,z_1)={\cal T}\left\{e^{-i\int_{z_1}^{z_2}dz'\hat{H}(z')}\right\},
\end{equation}
where the sign ${\cal T}$ stands for the contour time ordering. The result holds also for the case of time independent operator $\hat{O}$, but we keep the contour argument of the operator in order to specify its position on the contour, especially under the contour time ordering ${\cal T}$.

The n-particle Green's functions obey the boundary condition
\begin{equation}
G_n(1,\cdots,{\bf x}_k,z_{\text{i}},\cdots,n;1',\cdots,n')=G_n(1,\cdots,{\bf x}_k,z_{\text{f}},\cdots,n;1',\cdots,n'),
\end{equation}
known as the Kubo-Martin-Schwinger relation.

\begin{figure}
\includegraphics[width=0.5\linewidth]{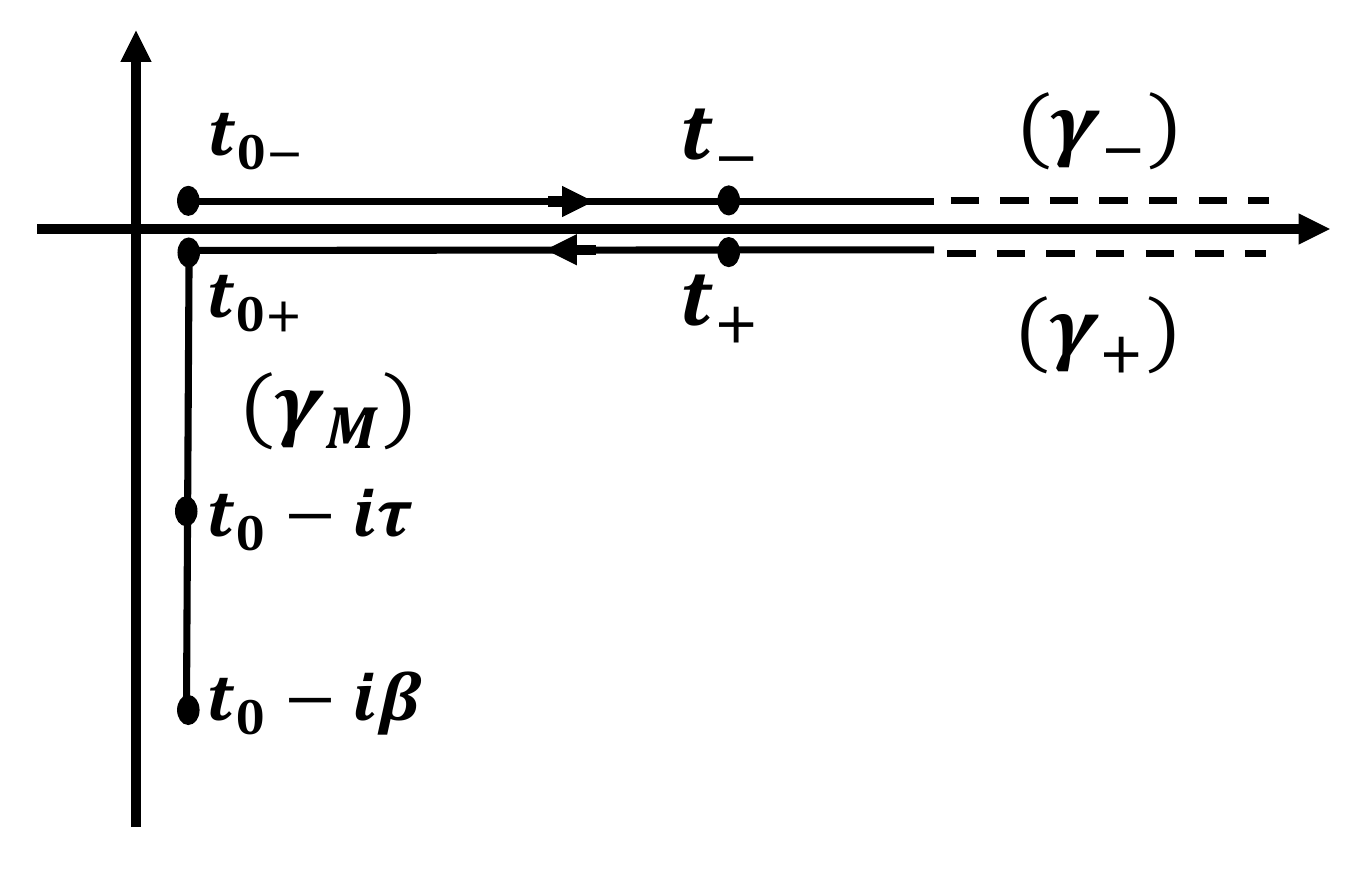}
\caption{The contour  $\gamma\equiv\gamma_-\oplus\gamma_+\oplus\gamma^M$ in the complex plane is presented. The horizontal part along the real axis contains two branches, the forward branch $\gamma_-$ from $t_0$ up to $t$, and the backward branch $\gamma_+$ from $t$ back to $t_0$, where the contour is extended to infinity. The vertical branch $\gamma_M$ is along the imaginary axis from $t_0$ to $t_0-i\beta$.}
\label{PhotPhonDis}
\end{figure}

On the horizontal part of the contour, if $z$ lies on the $\gamma_-$ and $\gamma_+$ branches,  in the n-particle Green's function Eq.(\ref{nGreen}) we get the time dependent quantum average. For a physical observable we have $\hat{O}(z'=t'_{\pm})\equiv\hat{O}(t')$, e.g. for the Hamiltonian we have $\hat{H}(z'=t'_{\pm})\equiv\hat{H}(t')$. For the field operators in the Schrodinger picture we have $\hat\psi({\bf x},z'=t_{\pm})\equiv \hat\psi({\bf x})$  and $\hat\psi^{\dagger}({\bf x},z'=t_{\pm})\equiv \hat\psi^{\dagger}({\bf x})$, and in the Heisenberg picture we have  $\hat\psi_H({\bf x},z)\rightarrow\hat\psi_H({\bf x},t)$ and $\hat\psi_H^{\dagger}({\bf x},z)\rightarrow\hat\psi_H^{\dagger}({\bf x},t)$. Moreover, concerning our previous Hamiltonian we have $h(z=t_{\pm})=h(t)$ and $v({\bf x},{\bf x}',z=t_{\pm})=v({\bf x},{\bf x}',t)$.

If $z$ lies on $\gamma_M$, the vertical part of the contour,  in the n-particle Green's function Eq.(\ref{nGreen}) we get the ensemble average at thermal equilibrium. Now $\hat{O}(z)$ is the same at every point, and we use $\hat{O}^M\equiv\hat{O}(z\in\gamma^M)$, where we choose $\hat{O}^M=\hat{O}(t_0)$. For the Hamiltonian we have $\hat{H}^M=\hat{H}(z\in\gamma^M)$, where $\hat{H}^M=\hat{H}(t_0)-\mu \hat{N}$, with the number operator $\hat{N}=\int d{\bf x}~\hat\psi^{\dagger}({\bf x})\hat\psi ({\bf x})$, and $\mu$ is the chemical potential. The field operators on the contour $\gamma_M$ are constant, then we have $\hat\psi({\bf x},z\in\gamma^M)\equiv\hat\psi({\bf x})$ and $\hat\psi^{\dagger}({\bf x},z\in\gamma^M)\equiv\hat\psi^{\dagger}({\bf x})$. In our case of the above Hamiltonian we have $h(z\in\gamma^M)=h^M$, where $h^M=h-\mu$, and $v({\bf x},{\bf x}',z\in\gamma^M)=v^M({\bf x},{\bf x}')$.

The one-particle Green's function, $G(1;1')\equiv G_1(1;1')$, equation of motion reads
\begin{equation}
\left[i\frac{d}{dz_1}-h(1)\right]G(1;1')=\delta(1;1')+ i\int d2~v(1;2)G_2(1,2;1',2^+),
\end{equation}
and the two-particle Green's function equation of motion is
\begin{eqnarray}
\left[i\frac{d}{dz_1}-h(1)\right]G_2(1,2;1',2')&=&\delta(1;1')G(2;2')\pm\delta(1;2')G(2;1') \\ \nonumber
&+&i\int d3~v(1;3)G_3(1,2,3;1',2',3^+),
\end{eqnarray}
and so on for higher order Green's functions. We get a system of coupled differential equations that is known as the Martin-Schwinger hierarchy for the Green's functions. The delta function is defined by $\delta(j;k)\equiv \delta(z_j;z_k)\delta({\bf x}_j-{\bf x}_k)$, where $\delta(z,z')$ is zero everywhere except in $z=z'$. On the contour $\gamma$ the $\delta$-function is zero if $z$ and $z'$ lie on different branches, namely $\delta(t_{\pm},t_{\mp})=0$. Due to the orientation of the contour we have $\delta(t_-,t'_-)=\delta(t-t')$ and $\delta(t_+,t'_+)=-\delta(t-t')$ where $\delta(t,t')$ is the real axis $\delta$-function. On the vertical branch we get $\delta(t_0-i\tau,t_0-i\tau')=i\delta(\tau-\tau')$. We have for the diagonal Hamiltonian $\langle {\bf x}_j|\hat{h}(z_j)|{\bf x}_k\rangle=\delta({\bf x}_j-{\bf x}_k)h(j)$, with $h(j)=h({\bf x}_j,-i{\bf \nabla}_j,z_j)$. Note that $h(z)=h^M=h-\mu$ on the vertical part for $z=t_0-i\tau$, and $h(z)=h$ on the horizontal branches for $z=t_{\pm}$. We introduced $v(j;k)\equiv\delta(z_j;z_k)v({\bf x}_j,{\bf x}_k,z_j)$. Explicitly we get $v({\bf x},z;{\bf x}',z')=\delta(z,z')v({\bf x},{\bf x}',t)$ on the horizontal branches of $\gamma$ with $z=t_{\pm}$, and $v({\bf x},z;{\bf x}',z')=\delta(z,z')v^M({\bf x},{\bf x}')$ on the vertical part of $\gamma$. Furthermore, in the notation $(j^+={\bf x}_j,z_j^+)$ the $z_j^+$ is infinitesimally later than $z_j$, (and in $(j^-={\bf x}_j,z_j^-)$ the $z_j^-$ is infinitesimally earlier than $z_j$). The analytical properties of two point correlators that belong to the Keldysh space is presented in the appendix.

\section{The T-Matrix approximation}

We obtain an infinite hierarchy of equations in which the equation of motion for $G_n$ is related to $G_{n+1}$ and $G_{n-1}$. The hierarchy can be truncated by appealing to appropriate approximations. We apply the T-matrix approximation that holds for short range interactions in the limit of low density particles. The two-particle Green's function can be written as
\begin{equation}
G_2(1,2;1',2')=G(1;1')G(2;2')\pm G(1;2')G(2;1')+\Gamma(1,2;1',2'),
\end{equation}
where the first two terms give the Hartree-Fock approximation, and the last term includes the $\Gamma$ correlation function. To the lowest order in the interaction, the solution for the correlation function yields
\begin{equation}
\Gamma(1,2;1',2')\approx i\int d3d4~G_0(1;3)G_0(2;4)v(3;4)G_2(3,4;1',2'),
\end{equation}
where the one-particle non-interacting Green's function, $G_0$, obeys $\left[i\frac{d}{dz_1}-h(1)\right]G_0(1;1')=\delta(1;1')$. The $G_2$ can be written as
\begin{equation}
G_2(1,2;1',2')=\int d3d4~S(1,2;3,4)\left[G(3;1')G(4;2')\pm G(3;2')G(4;1')\right],
\end{equation}
where we define the $S$ function by
\begin{equation}
S(1,2;3,4)=\delta(1;3)\delta(2;4)+i\int d5d6~G_0(1;5)G_0(2;6)v(5;6)S(5,6;3;4).
\end{equation}
The $T_0$ matrix is defined by $T_0(1,2;1',2')\equiv v(1;2)S(1,2;1',2')$. Multiplying $G_2$ by $v$ we get
\begin{equation}
v(1;2)G_2(1,2;1',2')=\int d3d4~T_0(1,2;3,4)\left[G(3;1')G(4;2')\pm G(3;2')G(4;1')\right].
\end{equation}
Using the above definition of $S$, we obtain
\begin{equation}
T_0(1,2;1',2')=\delta(1;1')\delta(2;2')v(1',2')+i\int d3d4~T_0(1,2;3,4)G_0(3;1')G_0(4;2')v(1';2').
\end{equation}
The T-matrix is a tool for studying the scattering of particles in quantum mechanics. Here we use it in order to study the scattering of particles in many-body problems, namely quantum scattering among particles in a medium of interacting particles \cite{Kadanoff1962}.

The T-matrix obeys the relation $T(1,2;1',2')=\delta(z_1,z_2)\delta(z'_1,z'_2)T({\bf x}_1,{\bf x}_2,z_1;{\bf x}'_1,{\bf x}'_2,z'_1)$, where we dropped the $0$ from the T-matrix, and in using the interaction property $v(j;k)=\delta(z_j,z_k)v({\bf x}_j,{\bf x}_k)$, we get
\begin{eqnarray}
&&T({\bf x}_1,{\bf x}_2,z_1;{\bf x}'_1,{\bf x}'_2,z'_1)=\delta(z_1,z'_1)\delta({\bf x}_1-{\bf x}'_1)\delta({\bf x}_2-{\bf x}'_2)v({\bf x}_1,{\bf x}_2) \\ \nonumber
&+&\int d{\bf x}_3d{\bf x}_4\int dz_3~T({\bf x}_1,{\bf x}_2,z_1;{\bf x}_3,{\bf x}_4,z_3){\cal G}_2({\bf x}_3,{\bf x}_4,z_3;{\bf x}'_1,{\bf x}'_2,z'_1)v({\bf x}'_1;{\bf x}'_2),
\end{eqnarray}
where ${\cal G}_2({\bf x}_3,{\bf x}_4,z_3;{\bf x}'_1,{\bf x}'_2,z'_1)=iG({\bf x}_3,z_3;{\bf x}'_1,z'_1)G({\bf x}_4,z_3;{\bf x}'_2,z'_1)$. Note that the one-particle Green's function is the non-interacting one $G_0$. At thermal equilibrium the functions depend only on the time difference, and the Fourier transform yields
\begin{eqnarray}
&&T({\bf x}_1,{\bf x}_2;{\bf x}'_1,{\bf x}'_2;\omega)=\delta({\bf x}_1-{\bf x}'_1)\delta({\bf x}_2-{\bf x}'_2)v({\bf x}_1,{\bf x}_2) \\ \nonumber
&+&\int d{\bf x}_3d{\bf x}_4~T({\bf x}_1,{\bf x}_2;{\bf x}_3,{\bf x}_4;\omega){\cal G}_2({\bf x}_3,{\bf x}_4;{\bf x}'_1,{\bf x}'_2;\omega)v({\bf x}'_1;{\bf x}'_2).
\end{eqnarray}

The T-matrix belong to the Keldysh space and contains a singular part (as presented in the appendix). We treat now  the retarded and advanced Keldysh components of the T-matrix. We have
\begin{equation}
T^{R/A}({\bf x}_1,{\bf x}_2;{\bf x}'_1,{\bf x}'_2;\omega)=\delta({\bf x}_1-{\bf x}'_1)\delta({\bf x}_2-{\bf x}'_2)v({\bf x}_1,{\bf x}_2)+\int \frac{d\omega'}{2\pi}\frac{{\cal B}({\bf x}_1,{\bf x}_2;{\bf x}'_1,{\bf x}'_2;\omega')}{\omega-\omega'\pm i\eta},
\end{equation}
where ${\cal B}({\bf x}_1,{\bf x}_2;{\bf x}'_1,{\bf x}'_2;\omega)=i\left[T^>({\bf x}_1,{\bf x}_2;{\bf x}'_1,{\bf x}'_2;\omega)-T^<({\bf x}_1,{\bf x}_2;{\bf x}'_1,{\bf x}'_2;\omega)\right]$. Moreover, we get the fluctuation-dissipation theorem $T^>({\bf x}_1,{\bf x}_2;{\bf x}'_1,{\bf x}'_2;\omega)=e^{\beta(\omega-2\mu)}T^<({\bf x}_1,{\bf x}_2;{\bf x}'_1,{\bf x}'_2;\omega)$. We obtain $T^<({\bf x}_1,{\bf x}_2;{\bf x}'_1,{\bf x}'_2;\omega)=-if(\omega-2\mu){\cal B}({\bf x}_1,{\bf x}_2;{\bf x}'_1,{\bf x}'_2;\omega)$ and $T^>({\bf x}_1,{\bf x}_2;{\bf x}'_1,{\bf x}'_2;\omega)=-i\bar{f}(\omega-2\mu){\cal B}({\bf x}_1,{\bf x}_2;{\bf x}'_1,{\bf x}'_2;\omega)$ with $f(\epsilon)=\frac{1}{e^{\beta\epsilon}-1}$ and $,\bar{f}(\epsilon)=1+f(\epsilon)$. In similar way, for ${\cal G}_2$ we get
\begin{equation}
{\cal G}_2^{R/A}({\bf x}_3,{\bf x}_4;{\bf x}'_1,{\bf x}'_2;\omega)=i\int \frac{d\omega'}{2\pi}\frac{{\cal G}_2^>({\bf x}_3,{\bf x}_4;{\bf x}'_1,{\bf x}'_2;\omega')-{\cal G}_2^<({\bf x}_3,{\bf x}_4;{\bf x}'_1,{\bf x}'_2;\omega')}{\omega-\omega'\pm i\eta}.
\end{equation}
The retarded ${\cal G}_2$ function is defined by
\begin{equation}
{\cal G}_2^R({\bf x}_3,{\bf x}_4;{\bf x}'_1,{\bf x}'_2;t_3-t'_1)=\theta(t_3-t'_1)\left\{{\cal G}_2^>({\bf x}_3,{\bf x}_4;{\bf x}'_1,{\bf x}'_2;t_3-t'_1)-{\cal G}_2^<({\bf x}_3,{\bf x}_4;{\bf x}'_1,{\bf x}'_2;t_3-t'_1)\right\},
\end{equation}
where as ${\cal G}_2^{\lessgtr}({\bf x}_3,{\bf x}_4;{\bf x}'_1,{\bf x}'_2;t_3-t'_1)=iG^{\lessgtr}({\bf x}_3;{\bf x}'_1;t_3-t'_1)G^{\lessgtr}({\bf x}_4;{\bf x}'_2;t_3-t'_1)$, the Fourier transform gives
\begin{equation}
{\cal G}_2^{R}({\bf x}_3,{\bf x}_4;{\bf x}'_1,{\bf x}'_2;\zeta)=i^2\int\frac{d\omega'}{2\pi}\frac{d\omega''}{2\pi}\frac{\left\{G^>({\bf x}_3;{\bf x}'_1;\omega')G^>({\bf x}_4;{\bf x}'_2;\omega'')-G^<({\bf x}_3;{\bf x}'_1;\omega')G^<({\bf x}_4;{\bf x}'_2;\omega'')\right\}}{\zeta-\omega'-\omega''},
\end{equation}
where $\zeta=\omega+i\eta$. The retarded T-matrix obeys
\begin{eqnarray}
&&T^R({\bf x}_1,{\bf x}_2;{\bf x}'_1,{\bf x}'_2;\zeta)=\delta({\bf x}_1-{\bf x}'_1)\delta({\bf x}_2-{\bf x}'_2)v({\bf x}_1,{\bf x}_2) \\ \nonumber
&+&\int d{\bf x}_3d{\bf x}_4~T^R({\bf x}_1,{\bf x}_2;{\bf x}_3,{\bf x}_4;\zeta){\cal G}_2^R({\bf x}_3,{\bf x}_4;{\bf x}'_1,{\bf x}'_2;\zeta)v({\bf x}'_1;{\bf x}'_2).
\end{eqnarray}
The retarded T-matrix is of big interest as it directly connected to the scattering amplitude.

\subsection{Momentum-Space Representation}

As a first step, due to transnational symmetry, we change variables and use the center of mass and relative positions, ${\bf X}=\frac{{\bf x}_1+{\bf x}_2}{2}$ and ${\bf x}={\bf x}_1-{\bf x}_2$, to get
\begin{eqnarray}
&&T^R({\bf x},{\bf X};{\bf x}',{\bf X}';\zeta)=\delta({\bf X}-{\bf X}')\delta({\bf x}-{\bf x}')v({\bf x}') \\ \nonumber
&+&\int d\bar{\bf x}d\bar{\bf X}~T^R({\bf x},{\bf X};\bar{\bf x},\bar{\bf X};\zeta){\cal G}_2^R(\bar{\bf x},\bar{\bf X};{\bf x}',{\bf X}';\zeta)v({\bf x}').
\end{eqnarray}
Now we define the center of mass and relative momentum by ${\bf P}={\bf p}_1+{\bf p}_2$ and ${\bf p}=\frac{{\bf p}_1-{\bf p}_2}{2}$, respectively. The Fourier transform and its inverse are defined by
\begin{equation}
{\cal O}({\bf x})=\int\frac{d{\bf p}}{(2\pi)^3}e^{i{\bf p}\cdot{\bf x}}{\cal O}({\bf p}),\ \ \ {\cal O}({\bf p})=\int d{\bf x}e^{-i{\bf p}\cdot{\bf x}}{\cal O}({\bf x}),
\end{equation}
and we use the identity $\delta({\bf x})=\int\frac{d{\bf p}}{(2\pi)^3}e^{
i{\bf p}\cdot{\bf x}}$ and its inverse $\delta({\bf p})=\int d{\bf x}e^{
-i{\bf p}\cdot{\bf x}}$. The Fourier transform of the T-matrix equation gives
\begin{equation}
T^R({\bf p};{\bf p}';{\bf P},\zeta)=v({\bf p}-{\bf p}')+\int \frac{d\bar{\bf p}}{(2\pi)^3}\frac{d\bar{\bf p}'}{(2\pi)^3}~T^R({\bf p};\bar{\bf p};{\bf P},\zeta){\cal G}_2^R(\bar{\bf p};\bar{\bf p}';{\bf P},\zeta)v(\bar{\bf p}'-{\bf p}').
\end{equation}
Here ${\bf p}$ is the initial momentum of one particle in the center of mass system, and ${\bf p}'$ is the final momentum of the scattered particle. The center of mass momentum is conserved, where the initial center of mass momentum ${\bf P}$ equals the final one ${\bf P'}$, (that is ${\bf P}={\bf P}'$). The momentum space coupling potential is
\begin{equation}
v({\bf p}-{\bf p}')=\int d{\bf r}'e^{-i({\bf p}-{\bf p}')\cdot{\bf x}'}v({\bf x}'),
\end{equation}
and the ${\cal G}_2^R$ is ${\cal G}_2^R(\bar{\bf p},\bar{\bf p}';{\bf P},\zeta)=\delta(\bar{\bf p}-\bar{\bf p}')\Upsilon(\bar{\bf p};{\bf P}\zeta)$, where
\begin{eqnarray}
\Upsilon(\bar{\bf p};{\bf P},\zeta)&=&i^2\int\frac{d\omega'}{2\pi}\frac{d\omega''}{2\pi}\left\{\frac{G^>(\frac{\bf P}{2}+\bar{\bf p};\omega')G^>(\frac{\bf P}{2}-\bar{\bf p};\omega'')}{\zeta-\omega'-\omega''}\right. \\ \nonumber
&-&\left.\frac{G^<(\frac{\bf P}{2}+\bar{\bf p};\omega')G^<(\frac{\bf P}{2}-\bar{\bf p};\omega'')}{\zeta-\omega'-\omega''}\right\}.
\end{eqnarray}
We have
\begin{equation}
T^R({\bf p};{\bf p}';{\bf P},\zeta)=v({\bf p}-{\bf p}')+\int \frac{d\bar{\bf p}}{(2\pi)^3}~T^R({\bf p};\bar{\bf p};{\bf P},\zeta)\Upsilon(\bar{\bf p};{\bf P},\zeta)v(\bar{\bf p}-{\bf p}').
\end{equation}

We assume local interaction, in which $v({\bf x})=v\delta({\bf x})$, and the Fourier transform gives $v({\bf p}-{\bf p}')=v$. Now, in the center of mass, the scattering is elastic where only the particle direction changes with $|{\bf p}|=|{\bf p}'|$. The T-matrix is depends only on the center of mass momentum, where $T^R({\bf p};{\bf p}';{\bf P},\zeta)\equiv T^R({\bf P},\zeta)$. We achieve
\begin{equation}
T^R({\bf P},\zeta)=\frac{v}{1-v\int \frac{d\bar{\bf p}}{(2\pi)^3}~\Upsilon(\bar{\bf p};{\bf P},\zeta)}.
\end{equation}
Let us assume a small center of mass momentum. We can consider ${\bf P}\approx 0$, where we have counter-propagating particles such that the total momentum is small, hence we have $T^R(\zeta)=\frac{v}{1-v\int \frac{d\bar{\bf p}}{(2\pi)^3}~\Upsilon(\bar{\bf p};\zeta)}$. The greater and lesser one-particle Green's functions are $G^<(\bar{\bf p};\omega)=- if(\omega){\cal A}(\bar{\bf p};\omega)$ and $G^>(\bar{\bf p};\omega)=- i\bar{f}(\omega){\cal A}(\bar{\bf p};\omega)$, respectively, where $f(\omega)=\frac{1}{e^{\beta(\omega-\mu)}-1}$ and $\bar{f}(\omega)=1+ f(\omega)=e^{\beta(\omega-\mu)}f(\omega)$. For non-interacting spectral function we have  ${\cal A}({\bf p};\omega)=2\pi\delta(\omega-\epsilon({\bf p}))$, where $\epsilon({\bf p})$ is the free particle dispersion. Using the dispersion symmetry $\epsilon({\bf p})=\epsilon(-{\bf p})$, we get $\Upsilon(\bar{\bf p};\zeta)=\frac{1+ 2f(\epsilon(\bar{\bf p}))}{\zeta-2\epsilon(\bar{\bf p})}$, and we obtain $1+ 2f(\epsilon(\bar{\bf p}))=\coth\left(\beta\frac{\epsilon({\bf p})-\mu}{2}\right)$. Finally we reach
\begin{equation}
T^R(\zeta)=\frac{v}{1-v\int \frac{d{\bf p}}{(2\pi)^3}~\frac{\coth\left(\beta\frac{\epsilon({\bf p})-\mu}{2}\right)}{\zeta-2\epsilon({\bf p})}}.
\end{equation}

\subsection{One-Dimensional Attractive Photons}

We consider now interacting photons propagating in one-dimensional nanoscale waveguide. The coupling parameter is negative for attractive interaction, where we make the change $v\rightarrow -v$, then in terms of the wavenumber
\begin{equation}
T^R(\zeta)=\frac{-v}{1+v\int \frac{dk}{2\pi}~\frac{\coth\left(\beta\frac{\epsilon(k)-\mu}{2}\right)}{\zeta-2\epsilon(k)}}.
\end{equation}
The linear dispersion is $\epsilon(k)=\epsilon_0+v_ek$, where $\epsilon_0$ is the minimum energy that appears due to transverse confinement, and $v_e$ is the photon effective group velocity. The interaction is limited to an energy band around the chemical potential where $|\epsilon(k)-\mu|<\Delta$. The band width is taken to be of the order of the phonon frequency $\Omega$. We change into the variable $\epsilon=\epsilon(k)-\mu$, where $d\epsilon=v_edk$, and we get
\begin{equation}
T^R(\zeta)=\frac{-v}{1+\frac{v}{2\pi v_e}\int_{-\Delta}^{+\Delta} d\epsilon~\frac{\coth\left(\beta\epsilon/2\right)}{\zeta-2\mu-2\epsilon}}.
\end{equation}
We evaluate the integral at the imaginary value of $\zeta-2\mu=i\eta$, where in the limit $\eta\rightarrow 0$ we obtain
\begin{equation}
T^R=\frac{-v}{1-g\int_{0}^{\Lambda}dx~\frac{\coth x}{x}},
\end{equation}
with $\Lambda=\beta\Delta/2$ and $g=\frac{v}{2\pi v_e}$, after making the change of variable of $x=\beta\epsilon/2$.

The T-matrix measures the probability amplitude for adding a pair of particles into the system and afterward removing a pair. A complex pole in the upper half plane indicates that if a pair of particles with opposite momenta are added at a certain time, the probability amplitude for removing such a pair increases exponentially in time, and the T-matrix approximation breakdown. The appearance of complex poles signals the formation of bound states (photon molecules). In the limit of high temperature $T\rightarrow\infty$, where $\beta\rightarrow 0$ and $x\rightarrow 0$, we have $1<g\int_{0}^{\Lambda}dx\frac{\coth x}{x}$, as $\coth x\rightarrow \infty$. Hence the T matrix contains no complex poles. On the other hand, in the limit of low temperature $T\rightarrow 0$, where $\beta\rightarrow \infty$ and $x\rightarrow \infty$, we have $1\ge g\int_{0}^{\Lambda}dx\frac{\coth x}{x}$, as $\coth x\rightarrow 1$. Hence at low temperature complex poles can appear in the T-matrix.

The critical temperature for the formation of bound states can be estimated from the equality $1=g\int_{0}^{\Lambda}dx\frac{\coth x}{x}$, when the denominator of the T-matrix vanishes. We are in the limit of $\Delta\ll\beta$, and the integral is taken in a region where $\coth x\approx 1$. As we are in the limit of low temperature one can integrate in the region of $x$ between $1$ and $\Lambda$, where $1\approx g\int_{1}^{\Lambda}dx\frac{1}{x}$. Then we get $1=g\ln \Lambda$, where we can write $\frac{2\pi v_e}{v}=\ln \frac{\Delta}{2k_BT_c}$, that lead to the result of 
\begin{equation}
k_BT_c=\frac{\Delta}{2}e^{-\frac{2\pi v_e}{v}}.
\end{equation}
We choose the numbers such that $v/v_e=4\pi$, and the band width is fixed by the vibrational mode frequency of about $\Omega=40$~GHz, with $\Delta=\hbar\Omega$ \cite{Zoubi2017}. Then we have $k_BT_c=\frac{\hbar}{\sqrt{e}}~20\text{GHz}$, where the critical temperature is $T_c\approx 0.1$~K.

\section{Photon Molecules}

After demonstrating the possibility of the formation of photon molecules in the previous section, we represent here the photon bound states and show how to implement them for quantum logic gates. We start with the three dimensional case and concentrate later in the one dimensional waveguide case. We consider two counter-propagating photons, $(a)$ and $(b)$, that form a molecule of wavevector ${\bf K}$ and is characterized by a normalized wavefunction $\chi({\bf x}_a-{\bf x}_b)$. The photon molecule wavefunction is defined  by
\begin{equation}
\phi_{\bf K}({\bf x}_a,{\bf x}_b)=\frac{1}{\sqrt{V}}~e^{i{\bf K}\cdot({\bf x}_a+{\bf x}_b)/2}\chi({\bf x}_a-{\bf x}_b),
\end{equation}
where $V=L^3$ is the system volume. Using transnational symmetry, then the Fourier transform reads $\tilde{\chi}_{\bf k}=\frac{1}{\sqrt{V}}\int d^3x~e^{-i{\bf k}\cdot{\bf x}}\chi({\bf x})$ with the inverse transform $\chi({\bf x})=\frac{1}{\sqrt{V}}\sum_{\bf k}\tilde{\chi}_{\bf k}~e^{i{\bf k}\cdot{\bf x}}$, to get
\begin{equation}
\phi_{\bf K}({\bf x}_a,{\bf x}_b)=\frac{1}{V}\sum_{\bf k}\tilde{\chi}_{\bf k}~e^{i\left(\frac{\bf K}{2}+{\bf k}\right)\cdot{\bf x}_a}e^{i\left(\frac{\bf K}{2}-{\bf k}\right)\cdot{\bf x}_b}.
\end{equation}
We use the periodic boundary condition, then the wavenumber ${\bf k}=(k_x,k_y,k_z)$ takes the values $k_i=2\pi n_i/L$ with $(i=x,y,z)$, and $n_i$ are integers. The normalization condition is given by $\int d^3x\left|\chi({\bf x})\right|^2=\sum_{\bf k}\left|\tilde{\chi}_{\bf k}\right|^2=1$.

The ket eigenstate is given by $|\phi_{\bf K}(a,b)\rangle=\sum_{\bf k}\tilde{\chi}_{\bf k}\left|\left.a,\left(\frac{\bf K}{2}+{\bf k}\right);b,\left(\frac{\bf K}{2}-{\bf k}\right)\right.\right\rangle$. The ket can be created from the vacuum using the second quantization operators of Fock's space, by $|\phi_{\bf K}\rangle=\sum_{\bf k}\tilde{\chi}_{\bf k}~\hat{a}_{\frac{\bf K}{2}+{\bf k}}^{\dagger}\hat{a}_{\frac{\bf K}{2}-{\bf k}}^{\dagger}|\text{vac}\rangle$. We define the pair creation operator of a molecule having a total wavenumber ${\bf K}$ by $\hat{A}_{\bf K}^{\dagger}=\sum_{\bf k}\tilde{\chi}_{\bf k}~\hat{a}_{\frac{\bf K}{2}+{\bf k}}^{\dagger}\hat{a}_{\frac{\bf K}{2}-{\bf k}}^{\dagger}$. The creation operator is related to the field operator by $\hat{a}^{\dagger}_{\bf k}=\frac{1}{\sqrt{V}}\int d^2x~e^{i{\bf k}\cdot{\bf x}}\hat{\psi}^{\dagger}({\bf x})$, which lead to
\begin{equation}
\hat{A}_{{\bf K}}^{\dagger}=\frac{1}{\sqrt{V}}\int d^3X~e^{i{\bf K}\cdot{\bf X}}\int d^3x~{\chi}({\bf x})~\hat{\psi}^{\dagger}({\bf X}+{\bf x}/2)\hat{\psi}^{\dagger}({\bf X}-{\bf x}/2).
\end{equation}
The pair field operator can be defined as $\hat{\Psi}^{\dagger}({\bf X})=\frac{1}{\sqrt{V}}\sum_{\bf K}e^{-i{\bf K}\cdot{\bf X}}\hat{A}_{{\bf K}}^{\dagger}$. We obtain, using the result $\sum_{\bf K}e^{i{\bf K}\cdot\left({\bf X'}-{\bf X}\right)}=V\delta({\bf X}-{\bf X'})$, the pair field creation operator
\begin{equation}
\hat{\Psi}^{\dagger}({\bf X})=\int d^3x~{\chi}({\bf x})~\hat{\psi}^{\dagger}({\bf X}+{\bf x}/2)\hat{\psi}^{\dagger}({\bf X}-{\bf x}/2).
\end{equation}

\subsection{Quantum Nonlinear Phase}

The wavefunction of a photon molecule build of two counter-propagating pair of photons can accumulate a quantum nonlinear phase of the order of $\pi$ in the appropriate condition. The phase is shown to be useful for the implementation of the photon molecule as a quantum logic gate. We consider two counter propagating photons, ($a$) and ($b$), of minimum frequency $\omega_0$, and with effective group velocity $v_e$, inside one-dimensional nanoscale wires. The free photon real space Hamiltonian is
\begin{equation}
\hat{H}_0=\int dx\left\{\omega_0\ \hat{\psi}_a^{\dagger}(x)\hat{\psi}_a(x)+iv_e\frac{\partial\hat{\psi}_a^{\dagger}(x)}{\partial x}\hat{\psi}_a(x)+\omega_0\ \hat{\psi}_b^{\dagger}(x)\hat{\psi}_b(x)-iv_e\frac{\partial\hat{\psi}_b^{\dagger}(x)}{\partial x}\hat{\psi}_b(x)\right\},
\end{equation}
and the photon-photon interaction Hamiltonian of coupling parameter $v$ is given by
\begin{equation}
\hat{H}_I=\frac{v}{2}\int dxdx'\ \delta(x-x')\ \hat{\psi}_a^{\dagger}(x')\hat{\psi}_b^{\dagger}(x)\hat{\psi}_a(x')\hat{\psi}_b(x).
\end{equation}
The photon pair  bound-state is given by
\begin{equation}
|\Psi(t)\rangle=\int d(x_1-x_2)\ \phi(x_1-x_2,t)\ \hat{\psi}_a^{\dagger}(x_1)\hat{\psi}_b^{\dagger}(x_2)|\text{vac}\rangle.
\end{equation}
The state obeys the Schrodinger equation $i\frac{\partial}{\partial t}|\Psi(t)\rangle=\hat{H}|\Psi(t)\rangle$, and we get
\begin{equation}
i\left\{\frac{\partial}{\partial t}+v_e\left(\frac{\partial}{\partial x_1}-\frac{\partial}{\partial x_2}\right)\right\}\phi(x_1-x_2,t)=\left\{2\omega_0+\frac{v}{2}\delta(x_1-x_2)\right\}\phi(x_1-x_2,t).
\end{equation}
Moving into rotating frame by using $\phi(x_1-x_2,t)=\tilde{\phi}(x_1-x_2,t)e^{-2i\omega_0t}$, gives
\begin{equation}
i\left\{\frac{\partial}{\partial t}+v_e\left(\frac{\partial}{\partial x_1}-\frac{\partial}{\partial x_2}\right)\right\}\tilde{\phi}(x_1-x_2,t)=\frac{v}{2}\delta(x_1-x_2)\tilde{\phi}(x_1-x_2,t).
\end{equation}
In term of the center of mass and relative positions, $X=\frac{x_1+x_2}{2}$ and $x=x_1-x_2$, where
\begin{equation}
\frac{\partial}{\partial X}=\frac{\partial}{\partial x_1}+\frac{\partial}{\partial x_2},\ \ \ 2\frac{\partial}{\partial x}=\frac{\partial}{\partial x_1}-\frac{\partial}{\partial x_2},
\end{equation}
we get
\begin{equation}
i\left\{\frac{\partial}{\partial t}+2v_e\frac{\partial}{\partial x}\right\}\tilde{\phi}(x,t)=\frac{v}{2}\delta(x)\tilde{\phi}(x,t).
\end{equation}
We apply the change of variables $\eta=x-2v_et$ and $\xi=x$, where
\begin{equation}
\frac{\partial}{\partial X}=\frac{\partial}{\partial\eta}+\frac{\partial}{\partial\xi},\ \ \ \frac{\partial}{\partial t}=-2v_e\frac{\partial}{\partial\eta},
\end{equation}
and we obtain $\frac{\partial}{\partial t}+2v_e\frac{\partial}{\partial x}=2v_e\frac{\partial}{\partial\xi}$, then the equation of motion
\begin{equation}
\frac{\partial}{\partial\xi}\tilde{\phi}(\eta,\xi)=-i\frac{v}{4v_e}\delta(\xi)\tilde{\phi}(\eta,\xi),
\end{equation}
and the solution reads
\begin{equation}
\tilde{\phi}(\eta,\xi)=\tilde{\phi}_{\text{in}}e^{-i\frac{v}{4v_e}\int d\xi\delta(\xi)}=\tilde{\phi}_{\text{in}}e^{-i\frac{v}{4v_e}}.
\end{equation}

\begin{table}[ht]
\caption{Z-Controlled Gate}
\centering
\begin{tabular}{c c}
\hline\hline
Input State$\ \ \ $ & $\ \ \ $Output State  \\ [0.5ex]
\hline
$|0,0\rangle$  &  $\ |0,0\rangle$  \\
 $|1,0\rangle$  &  $\ |1,0\rangle$  \\
 $|0,1\rangle$  &  $\ |0,1\rangle$  \\
 $|1,1\rangle$  & -$|1,1\rangle$  \\ [1ex]
\hline
\end{tabular}
\label{table:nonlin}
\end{table}

Photon molecules can serve as a tool for the implementation of quantum logic gates. We show how to achieve Z-controlled gate for the case of $\vartheta=\pi$ where the nonlinear quantum phase is $\vartheta=\frac{v}{4v_e}$, which is obtained for the previous value of $v/v_e=4\pi$. If in the input we have zero photons at the two channels, that is $|0,0\rangle$, the output is the same state of $|0,0\rangle$. If in the input one channel includes one photon and the second is empty, that is $|0,1\rangle$ or $|1,0\rangle$, then the output is again the same state, $|0,1\rangle$ or $|1,0\rangle$. While for the case of two counter-propagating photons, then the input state is $|1,1\rangle$, and now the output state acquires a phase where $-|1,1\rangle$, as presented in table (1). The gate is universal and in combination with photon Hadamard gates one can achieve all requested quantum logic gates.

\section{Conclusions}

Quantum nonlinear optics is a hot topic and found big interest in the recent years for its importance to both fundamental physics and applications in quantum information processing. Several proposals have been suggested recently for the realization of quantum nonlinear optics and mainly for achieving strongly interacting photons, while for each proposal exists the advantages and disadvantages. Nanoscale structures are solid components and then can be easily integrated into all-optical platforms, the fact that makes them very attractive. Many-body physics of photons can be treated in using different techniques that rest on approximations. In the present paper we adopted the method of contour Green's functions that permits one to extract the system physical properties. We reached a hierarchy of equations of motion for the Green's functions, where the equation for one-particle Green's function depends on the equation for two-particle Green's function and so on for higher orders. The system is unsolvable exactly and we applied the T-matrix approximation that allows the truncation of the equations, and that can be solved for the scattering problem in a medium of interacting particles.

In the paper we considered interacting slow photons in nanoscale wires, where we used our previously derived Hamiltonian for the effective photon-photon interaction that is mediated by phonons. We calculated the T-matrix and found the complex pole at which the approximation breaks down. The singularity in the T-matrix is the signature for the formation of photon bound states and provides the critical temperature at which such phenomena can appear. The photon bound state is represented by a ket state that is defined through two-particle creation operators. The equation of motion for the state amplitude is solved and results in nonlinear quantum phase shift that can be of the order of $\pi$ in the appropriate region of physical numbers. The photon molecules are shown to act as quantum logic gates by exploiting the nonlinear phase shift.

The present work can be extended into more interesting effects of many-body physics of photons in nanostructures. The contour Green's function is a strong tool that allows us to study further interesting phenomena, e.g. Bose-Einstein condensation and superfluidity of photons. Moreover, interacting photons in nanoscale structures is an ideal environment for studying non-equilibrium behavior, and the contour Green's function is the optimal tool for achieving this aim. The implementation of interacting photons for quantum information processing is an important step toward introducing further nanophotonic quantum information components. Our setup allows us to achieve quantum information processing and communication in using hybridized components involving interacting photons. The issue we presented here can be extended into a wide range of quantum computing devices of photons.

\section*{Appendix}

We introduce here some analytical properties of the the general two point correlator that belongs to the Keldysh space. The two-point correlator is defined by ${\cal C}(z,z')=\text{Tr}\left[\hat{\rho}{\cal T}\left\{\hat{O}_1(z)\hat{O}_2(z')\right\}\right]$ for the two operators $\hat{O}_1(z)$ and $\hat{O}_2(z')$, where the matrix operator is given by $\hat\rho=\frac{e^{-\beta \hat{H}^M}}{Z}$, with the partition function $Z=\text{Tr}\left\{e^{-\beta \hat{H}^M}\right\}$. In Keldysh space we can write the general form
\begin{equation}
{\cal C}(z,z')={\cal C}^{\delta}\delta(z,z')+\theta(z,z'){\cal C}^>(z,z')+\theta(z',z){\cal C}^<(z,z'),
\end{equation}
where ${\cal C}^{\delta}$ is the singular part, and ${\cal C}^>(z,z')=\text{Tr}\left[\hat{\rho}\hat{O}_1(z)\hat{O}_2(z')\right]$, with ${\cal C}^<(z,z')=\text{Tr}\left[\hat{\rho}\hat{O}_2(z')\hat{O}_1(z)\right]$. On the backward and forward branches, as $\hat{O}_i(t_+)=\hat{O}_i(t_-)$ for $(i=1,2)$, we have ${\cal C}^{\lessgtr}(t_+,z')={\cal C}^{\lessgtr}(t_-,z')$, and ${\cal C}^{\lessgtr}(z,t'_+)={\cal C}^{\lessgtr}(z,t'_-)$. Also we have ${\cal C}^{\delta}(t_+)={\cal C}^{\delta}(t_-)\equiv {\cal C}^{\delta}(t)$. In general, we have $\theta(z,z')=1$ for $z$ later than $z'$ on the contour, and zero otherwise. Moreover, we have $\delta(z,z')=\frac{d}{dz}\theta(z,z')=-\frac{d}{dz'}\theta(z,z')$. We now define different Keldysh components that are functions of real variables. When both arguments on the horizontal branches, the greater and lesser Keldysh components are ${\cal C}^>(t,t')\equiv {\cal C}(t_+,t'_-)$, and ${\cal C}^<(t,t')\equiv {\cal C}(t_-,t'_+)$. Namely, the value of the contour function ${\cal C}^{\lessgtr}(z,z')$ is the real-time function ${\cal C}^{\lessgtr}(t,t')$. Furthermore, we define the retarded and advanced Keldysh components of real time arguments
\begin{eqnarray}
{\cal C}^R(t,t')&\equiv&{\cal C}^{\delta}\delta(t-t')+\theta(t-t')\left[{\cal C}^>(t,t')-{\cal C}^<(t,t')\right], \\ \nonumber
{\cal C}^A(t,t')&\equiv&{\cal C}^{\delta}\delta(t-t')-\theta(t'-t)\left[{\cal C}^>(t,t')-{\cal C}^<(t,t')\right].
\end{eqnarray}
The retarded component vanishes for $t<t'$, while the advanced component vanishes for $t>t'$. Dropping the singular part, the Fourier transform of the correlator, using ${\cal O}(t_1-t_2)=\int\frac{d\omega}{2\pi}e^{-i\omega(t_1-t_2)}{\cal O}(\omega)$ and the Heaviside identity $\theta(t_1-t_2)=i\int\frac{d\omega}{2\pi}\frac{e^{-i\omega(t_1-t_2)}}{\omega+i\eta}$, gives
\begin{equation}
{\cal C}^{R/A}(\omega)=\int\frac{d\omega'}{2\pi}\frac{\hat{\cal A}(\omega')}{\omega-\omega'\pm i\eta},
\end{equation}
where we defined the spectral function $\hat{\cal A}(\omega)=i\left[{\cal C}^>(\omega)-{\cal C}^<(\omega)\right]$ or $\hat{\cal A}(\omega)=i\left[{\cal C}^R(\omega)-{\cal C}^A(\omega)\right]$. We get the fluctuation-dissipation theorem ${\cal C}^<(\omega)=- i f(\omega-\mu)\hat{\cal A}(\omega)$ and ${\cal C}^>(\omega)=-i\bar{f}(\omega-\mu)\hat{\cal A}(\omega)$, where for bosons $f(\omega)=\frac{1}{e^{\beta\omega}+ 1}$ and $\bar{f}(\omega)=1\pm f(\omega)=e^{\beta\omega}f(\omega)$, with the relation ${\cal C}^>(\omega)=e^{\beta(\omega-\mu)}{\cal C}^<(\omega)$.

\section*{Acknowledgment}

The work was supported by the Council for Higher Education in Israel via the Maa'of Fellowship.


%

\end{document}